\documentclass[useAMS,usenatbib,usegraphicx]{mn2e}
\usepackage{times}
\usepackage{rotating}
\begin{document}

\def\cmma{\;\;\; ,}
\def\intl{\int\limits}
\def\ltk{\left [ \,}
\def\ltp{\left ( \,}
\def\ltb{\left \{ \,}
\def\rtk{\, \right  ] }
\def\rtp{\, \right  ) }
\def\rtb{\, \right \} }
\newcommand\jeuvb{$J_\nu^{\rm EUVB}$}
\newcommand\gell{\gamma_\ell}
\newcommand\mzq{z_{\rm em}}
\newcommand\zq{$\mzq$}
\newcommand\nqso{53}
\newcommand\vstep{0.025}
\newcommand\rmfpval{147 \pm 15} 
\newcommand\omfpval{242 \pm 42} 
\newcommand{\mhmpc}{\rm h_{70}^{-1} \, Mpc}
\newcommand{\hmpc}{$\mhmpc$}
\newcommand{\Lya}{Ly$\alpha$}
\newcommand{\lya}{Ly$\alpha$}
\newcommand{\lyb}{Ly$\beta$}
\newcommand{\mfnhi}{f(\mnhi)}
\newcommand{\fnhi}{$\mfnhi$}
\newcommand{\mflmfp}{\lambda_{\rm mfp,z=4}^{912}}
\newcommand{\mlmfp}{\lambda_{\rm mfp}^{912}}
\newcommand{\lmfp}{$\lambda_{\rm mfp}^{912}$}
\newcommand{\mteff}{\tau_{\rm eff,LL}}
\newcommand{\teff}{$\mteff$}
\newcommand{\mnhi}{N_{\rm HI}}
\newcommand{\nhi}{$\mnhi$}
\def\cm#1{\, {\rm cm^{#1}}}
\newcommand{\mtll}{\tau_{\rm 912}}
\newcommand{\tll}{$\mtll$}
\newcommand{\mdat}{{\delta \alpha_{\rm T}}}
\newcommand{\dat}{$\mdat$}
\newcommand{\mkppo}{\tilde \kappa_{912}(\mzq)}
\newcommand{\kppo}{$\mkppo$}
\newcommand{\mgigm}{\gamma_{\rm IGM}}
\newcommand{\gigm}{$\mgigm$}
\newcommand{\lox}{$\ell(X)$}
\newcommand{\loz}{$\ell(z)$}
\newcommand{\mnminr}{N_{\rm min}^{\rm R13}}
\newcommand{\nminr}{$\mnminr$}
\newcommand{\mtlya}{\tau_{\rm eff}^{\rm Ly\alpha}}
\newcommand{\tlya}{$\mtlya$}
\newcommand{\mtlyman}{\tau_{\rm eff}^{\rm Lyman}}
\newcommand{\tlyman}{$\mtlyman$}
\newcommand{\mkms}{{\rm km~s^{-1}}}
\newcommand{\tlox}{$\ell(X)_{\tau \ge 2}$}
\newcommand{\mztkll}{{\tilde\kappa}_{912}}
\newcommand{\ztkll}{$\mztkll$}
\newcommand{\lesssim}{\la}
\newcommand{\gtrsim}{\ga}
\def\ion#1#2{{#1}\,{\sevensize {#2}}}
\newcommand{\oneskip}{\vskip \baselineskip}
\newcommand{\annrev}{Annual Review of Astronomy \& Astrophysics}
\newcommand{\araa}{Annual Review of Astronomy \& Astrophysics}
\def\aap{Astronomy \& Astrophysics}
\def\aj{Astronomical Journal}
\def\apj{Astrophysical Journal}
\def\apjl{Astrophysical Journal Letters}
\def\apjs{Astrophysical Journal Supplements}
\def\mnras{Monthly Noticies of the Royal Astronomical Society}
\def\nat{Nature}
\def\pasp{Publications of the Astronomical Society of the Pacific}
\def\prd{Physical Reviews D}
\def\jcap{Journal of Cosmology and Astroparticle Physics}

\title[Unified IGM]{Towards a Unified Description of the Intergalactic
  Medium at Redshift $z \approx 2.5$} 

\author[Prochaska et al.]{
J. Xavier Prochaska$^{1,2}$,
Piero Madau$^1$,
John M. O'Meara$^3$
Michele Fumagalli$^{4,5,6}$ \\
$^1$Department of Astronomy and Astrophysics, University
  of California, 1156 High Street, Santa Cruz, CA 95064 USA. \\ 
$^2$University of California Observatories, Lick Observatory 
  1156 High Street, Santa Cruz, CA 95064 USA. \\ 
$^3$Department of Chemistry and Physics, Saint Michael's College.
One Winooski Park, Colchester, VT 05439 USA.\\
$^4$Carnegie Observatories, 813 Santa Barbara Street, Pasadena, CA 91101, USA. \\
$^{5}$Department of Astrophysics, Princeton University, Princeton, NJ 08544-1001, USA.\\
$^{6}$Hubble Fellow.\\
}

\maketitle
\begin{abstract}
We examine recent measurements of the $z \approx 2.5$ intergalactic
medium (IGM) which constrain the \ion{H}{I} frequency distribution
\fnhi\ and the mean free path \lmfp\ to ionizing radiation.  We argue
that line-blending and the clustering of strong absorption-line
systems have led previous authors to systematically overestimate the
effective Lyman limit opacity, yielding too small 
of a \lmfp\ for the IGM.  We further show that
recently published measurements of \fnhi\ at $\mnhi \approx 10^{16}
\cm{-2}$ lie in strong disagreement, implying underestimated
uncertainty from sample variance and/or systematics like
line-saturation. Allowing for a
larger uncertainty in the \fnhi\ measurements, we provide a new cubic
Hermite spline model for \fnhi\
which reasonably satisfies all of the observational constraints
under the assumption of randomly distributed absorption systems.  We
caution, however, that this formalism is invalid in light of absorber
clustering and use a toy model to estimate the effects.
Future work must properly account for the non-Poissonian nature of the
IGM.  
\end{abstract}


\section{Introduction}

The intergalactic medium (IGM, revealed by the \lya\ forest) is the diffuse medium of gas and
metals which traces the large-scale density fluctuations of the
universe.  These fluctuations were imprinted in primordial
density perturbations and, therefore,  
their analysis offers
unique constraints on cosmology, 
especially on scales of tens to hundreds of Mpc.
Modern observations of the IGM  --
via absorption-line analysis of distant quasars --
has provided several
probes of the $\Lambda$CDM paradigm including: 
a measurement of its
matter power spectrum, upper limits to the 
mass of neutrinos, and an independent measure of baryonic
acoustic oscillations \citep[e.g.][]{mcdonald05,vbh09,sik+13}. 

These analyses leverage the statistical power of
high-dispersion, high signal-to-noise (S/N)
spectra from a select set of sightlines
\citep[e.g.][]{cwb+03,bergeron04,rudie12},
together with low-dispersion, lower S/N
spectra on many thousands of sightlines
\citep{schneider+10,paris+12,lee+13}.
Concurrently, these datasets provide a precise
characterization of fundamental properties of the IGM.  This includes 
statistics on the opacity of the \lya\ forest
\citep[e.g.][]{cwb+03,fpl+08,palanque+13,bhw+13},
the incidence of optically thick gas 
\citep[aka Lyman limit systems or LLS,][hereafter O13]{pow10,ribaudo11,fop+13,omeara13},
and the \ion{H}{I} column densities (\nhi)
of the strongest, damped \lya\ systems 
\citep[DLAs;][]{phw05,pw09,noterdaeme+12}. 
Traditionally, many of these results have been described by a single
distribution function \fnhi\ defined at a given redshift and
often normalized to the absorption length
$dX = \frac{H_0}{H(z)}(1+z)^2 dz$ introduced by \cite{bp69}.

In principle, \fnhi\ encodes the primary characteristics of the IGM
and its evolution with cosmic time.  This includes an estimation of
the mean free path to ionizing radiation \lmfp, defined as the most
likely proper
distance a packet of ionizing photons will travel before suffering an
$\rm e^{-1}$ attenuation.  Under the standard assumption of randomly
distributed absorbers, 
one may calculate the effective Lyman limit opacity \teff\ as follows.
An ionizing photon with $h\nu_{\rm em} \ge 1$Ryd emitted from 
a quasar with redshift \zq\ will redshift to 1\,Ryd 
at $z_{912} \equiv (\nu_{\rm 912}/\nu_{\rm em})(1+\mzq) -1 $.
The effective optical depth that this photon experiences 
from Lyman limit continuum opacity
is then \citep[cf.][]{mm93}:

\begin{equation}
\mteff(z_{912},\mzq) = \intl_{z_{912}}^{\mzq} \intl_0^\infty f(\mnhi,z')
   \lbrace 1 - \exp \ltk - \mnhi \sigma_{\rm ph}(z') \rtk \rbrace
   d\mnhi dz' \;\;\; , 
\label{eqn:teff}
\end{equation}
where $\sigma_{\rm ph}$ is the photoionization cross-section evaluated
at the photon frequency
$\nu = \nu_{912}(1+z')/(1+z_{912})$.  For a given \zq,  
one may measure \lmfp\ by solving for
the redshift $z_{912}$ where $\mteff=1$ and converting the offset from
\zq\ to a proper distance.

In a series of recent papers, we have introduced an alternate method
to estimating \lmfp\ using composite quasar spectra,
which directly assess the average IGM
opacity to ionizing photons 
\citep{pwo09,omeara13,fop+13,worseck13}.
These `stacks' reveal the average, intrinsic quasar spectrum (its
spectral energy distribution or SED)
as attenuated by the IGM.
Provided one properly accounts for several other, secondary effects
on the composite spectrum at rest wavelengths $\lambda_r < 912$\AA\
(e.g.\ the underlying slope of the quasar SED), it is straightforward
to directly measure \lmfp\ and
estimate uncertainties from standard bootstrap techniques.
One may then use estimates of \lmfp\ to independently constrain \fnhi\ via
Equation~\ref{eqn:teff}, especially at column densities where \fnhi\ is
most difficult to estimate directly, i.e.\ at $\mnhi
\approx 10^{15} - 10^{17} \cm{-2}$ where the Lyman series lines lie on
the flat portion of the curve-of-growth and the Lyman limit opacity
$\mtll \ll 1$.
In this manner, we concluded that \fnhi\ is relatively flat 
at $\mnhi \lesssim 10^{17.5} \cm{-2}$ and steepens at lower column
densities \citep[][see also Ribaudo et al.\ 2011]{pow10,omeara13}.
These constraints may be compared against predictions from numerical
simulations to constrain models of galaxy formation and astrophsyical
processes related to radiative transfer
\citep[e.g.][]{mcquinn11,altay13}.

Most recently, \citet[][hereafter R13]{rudie13} have published a study on \fnhi\ for
$\mnhi \approx 10^{12} - 10^{17} \cm{-2}$ using the traditional approach of
performing Voigt profile fits to ``lines'' in quasar spectra. From
their unique dataset,
in terms of S/N and spectral coverage,  they report a high incidence of lines
with $\mnhi \approx 10^{16} \cm{-2}$. 
Using the incidence of LLS to extrapolate their results to higher
\nhi, 
R13 infer a much smaller mean free path at
$z=2.4$ ($\rmfpval \, \mhmpc$) than the direct measurement
($\omfpval \, \mhmpc$; O13).  Such a disagreement is unseemly in
this era of precision cosmology and IGM characterization.

More importantly, the 
difference has significant implications for
estimations of the intensity \jeuvb\ of 
the extragalactic ultraviolet background (EUVB),
the escape fraction from galaxies,
studies of \ion{He}{II} reionization, and
models of the circumgalactic medium of galaxies
\citep[e.g.][]{fumagalli11a,ccp13,nsk+13,dfm13,bb13}. 
For example,
a favored approach to evaluating \jeuvb\ is to calculate the
attenuation of known sources of ionizing radiation by the IGM
\citep{hm96,flh+08,hm12}.
A difference in \lmfp\ of a factor of 2 leads directly to a 100\%\
uncertainty in \jeuvb.  Similarly, a much lower \lmfp\ value yields a
systematically higher escape fraction from medium-band imaging below
the Lyman limit \citep[e.g.][]{nsk+13}.

In the following, we explore this apparent conflict and propose
several explanations to reconcile the measurements.  Furthermore, we 
offer new insight into the meaning and limitations
of \fnhi\ and its validity as a
description of the IGM.  Throughout the manuscript, we adopt a
$\Lambda$CDM cosmology with $\Omega_\Lambda = 0.7$, $\Omega_m=0.3$ and
$H_0 = 70 h_{70} \,\rm km \, s^{-1} \, Mpc^{-1}$ and we have translated
previous measurements to this cosmology where necessary.

\section{Controversies in the $z \approx 2.5$ IGM}
\label{sec:conflict}

In this section, we examine the primary observational constraints on
\fnhi\ and \lmfp\ in the $z \approx 2.5$ IGM and highlight tension
between the measurements.

\begin{figure}
\includegraphics[width=3.5in,angle=90]{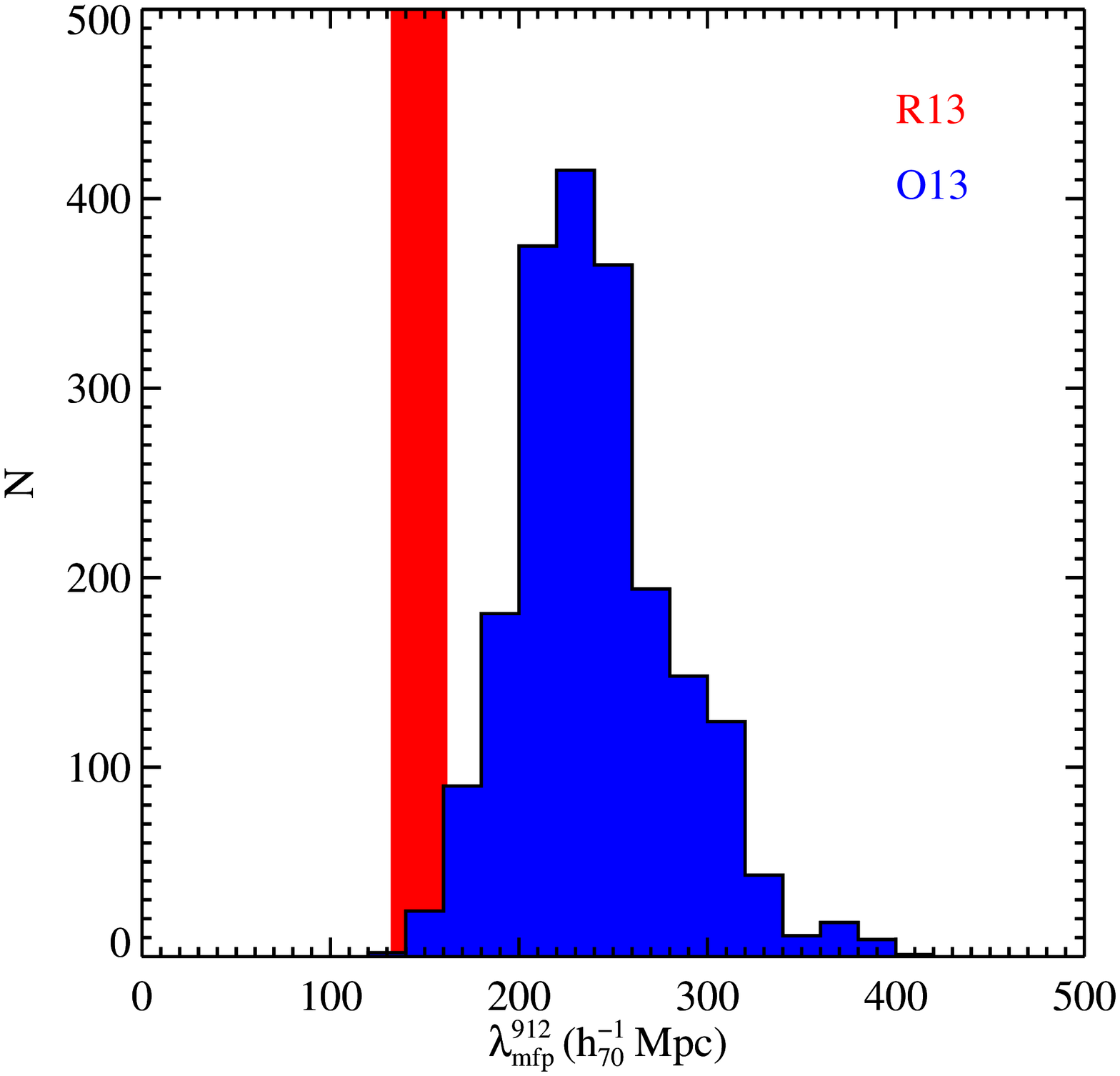}
\caption{The blue histogram shows the distribution of \lmfp\ values
  measured from the O13 quasar composite spectrum, adopting their
  bootstrap analysis and a $\Lambda$CDM cosmology with $\Omega_\Lambda = 0.7$,
  $\Omega_m=0.3$ and $H_0 = 70 h_{70} \,\rm km \, s^{-1} \, Mpc^{-1}$. 
  We measure a mean value $\mlmfp = \omfpval \mhmpc$.
  The red band shows the R13 estimate of \lmfp: $\rmfpval \mhmpc$.
  Only 31/2000 of the trials fall within $1\sigma$ of the R13 value.
}
\label{fig:compare_mfp}
\end{figure}

\subsection{Comparison of the Published MFP Values}

There are two recently published values for the MFP to
ionizing photons at $z \approx 2.4$, derived from two distinct
techniques: 
(1) an evaluation based on the measured, average attenuation of 
flux in a composite quasar spectrum ($\omfpval \, \mhmpc$; O13); 
and
(2) the value derived by R13 ($\rmfpval \, \mhmpc$) 
from their new constraints on \fnhi, old
estimations of the incidence of LLS, and an assumed redshift evolution
in the line density of strong absorbers $\ell(z) \propto
(1+z)^{\gell}$.  
R13 fitted a disjoint, double power-law model to the
\fnhi\ constraints, generated IGM models with standard Monte Carlo
techniques, and assessed the predicted flux attenuation to estimate\footnote{ 
  We have confirmed that the central \lmfp\ value of R13 matches
  that recovered from the evaluation of Equation~\ref{eqn:teff} with
  their favored \fnhi\ model.}
\lmfp\ and its uncertainty.  
Treating the uncertainty in each measurement as a Gaussian, the two
diverge at 97\%c.l.
The uncertainty in \lmfp\ from O13 is non-Gaussian, however, with a
significant tail to higher values.
Figure~\ref{fig:compare_mfp} shows 2000\,bootstrap evaluations of
\lmfp\ from O13, rerun with the cosmology adopted here.  The average
value is $\omfpval \, \mhmpc$, and we find that only 31/2000 trials have
\lmfp\ within $1\sigma$ of the R13 value (shaded region).

Worseck et al.\ (2013) have recently analyzed the redshift evolution
of \lmfp\ combining all of the $z>2$ measurements made from quasar
composite spectra.  They find the values are well modeled by a single
power-law $\mlmfp(z) = \mflmfp [(1+z)/5]^\eta$ with $\mflmfp = 35
\, \mhmpc$ and $\eta = -5.45$. This gives $\mlmfp = 244 \mhmpc$ at
$z=2.5$, exceeding the R13 value at very high confidence.
Similarly, previous
estimations based on evolution in the incidence of LLS 
\citep[e.g.][]{songaila10} yields larger values than that reported
by R13, as
illustrated by R13 (their Figure~15).
We conclude that the O13 and R13 measurements of \lmfp\ at $z \approx 2.5$
are highly discrepant.
We do note that the quasars sampled by R13 have emission redshifts 0.2
to 0.3 higher than the quasars of O13 and also systematically large
luminosities.  We do not believe, however, that this drives any of the
differences in the derived IGM properties.

\subsection{Modeling the Quasar Composite Spectrum}

We further examine the tension between O13 and R13 with
the methodology used by O13 to measure \lmfp.
Those authors fit a $z \approx 2.5$ 
quasar composite spectrum at rest wavelengths
$\lambda_r < 912$\AA\ with a five parameter model:
 (i) a power-law tilt to the assumed Telfer quasar template
 \citep{telfer02}, \dat;
 (ii) the normalization of the quasar SED at $\lambda_r = 912$\AA,
 $C_{\rm T}$; 
 (iii) the redshift evolution of the integrated Lyman series opacity,
 $\gamma_\tau$; 
 and
 (iv,v) the normalization and redshift evolution of \lmfp\ which O13
 parameterized in terms of an opacity:

\begin{equation}
\mztkll(z) = \mkppo \ltk \frac{1+z}{1+\mzq}
\rtk^{\gamma_\kappa} \;\;\; .
\label{eqn:kLL_z}
\end{equation}
The resultant values of \kppo\ and $\gamma_\kappa$ provide an estimate
for the mean free path (Figure~\ref{fig:compare_mfp}).  

\begin{figure}
\includegraphics[width=3.0in,angle=90]{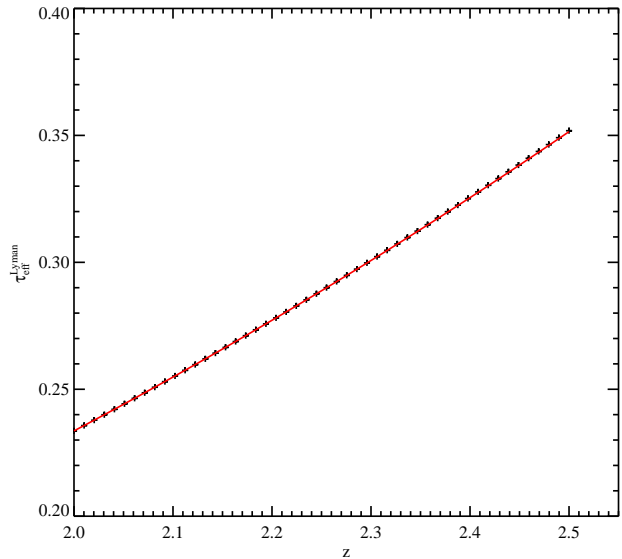}
\caption{Plot of the predicted effective optical depth of the full
  \ion{H}{I} Lyman series \tlyman\ assuming the \fnhi\ model of R13
  and a uniform Doppler parameter $b=24 \,\mkms$. 
  The red curve is an assumed power-law description of the values, 
  $\mtlyman(z) = \tau_{\rm eff, z=2}^{\rm Lyman}
  [(1+z)/3]^{\gamma_{\tau}}$ with $\gamma_\tau = 2.65$.
}
\label{fig:gtau}
\end{figure}

We can use the same methodology to find the best-fitting model
of the quasar composite spectrum,
constrained\footnote{ Achieved in practice
by restricting the combined values of \kppo\ and $\gamma_\kappa$.} 
to yield the \lmfp\ value reported by R13. 
This model adopts the same redshift evolution of the frequency
distribution of the IGM assumed by R13.  Specifically,
R13 assumed $\gell = 1$ for the redshift evolution in the incidence of 
strong absorbers
($\mnhi > 10^{15.1} \cm{-2}$) and $\gell = 2.5$ for lower
\nhi\ lines.  Empirically, we find that 
this implies $\gamma_\tau \approx 2.65$ (Figure~\ref{fig:gtau}).
Finally, 
the parameter \dat\ was allowed to freely vary between $[-0.8,0]$ and
we let $C_{\rm T}$ vary by $\pm 10\%$, following O13.

\begin{figure}
\includegraphics[width=3.1in,angle=90]{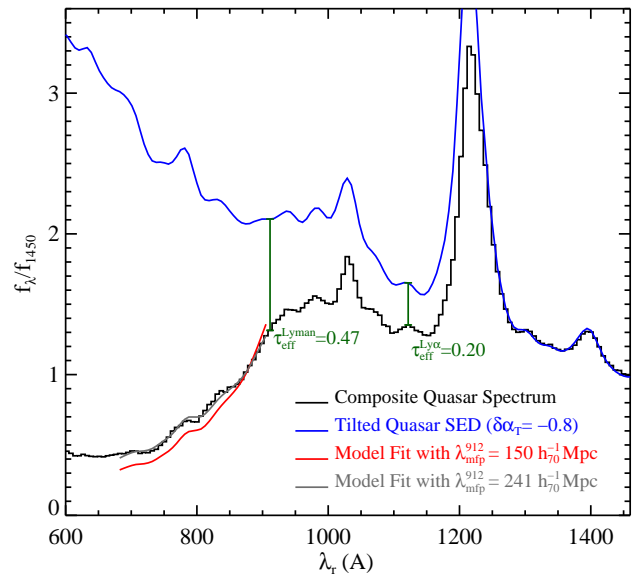}
\caption{The black curve is the composite quasar spectrum constructed
  by O13 from their HST/WFC3 observations of 53 quasars at $z \approx
  2.5$.  Overplotted at $\lambda_r = 680-910$\AA\ as a gray curve
  is the best-fit
  model of O13 which corresponds to a mean free path $\mlmfp = 242
  \, \mhmpc$.  The red curve is the best-fit model forced to yield the
  $\mlmfp = 150 \, \mhmpc$ value of R13.  It is a very poor description of the data.
  The blue curve is the intrinsic quasar SED for the $\mlmfp = 150 \,
  \mhmpc$ model, 
  restricted to match the data at 1450\AA.
  It is the Telfer quasar template \citep{telfer02} ``tilted'' by
  $\mdat = -0.8$ with $f^{\rm SED}_\lambda = f_\lambda^{\rm Telfer}
  (\lambda/1450 \, {\rm \AA})^\mdat$.  This quasar SED yields an
  acceptable measurement for the effective \lya\ opacity \tlya\ at
  $\lambda_r \approx 1100$\AA\ but exceeds the effective Lyman series
  opacity \tlyman\ at $\lambda_r = 912$\AA\ predicted by R13.  
  A quasar SED with $\mdat = -2$ would yield an acceptable model of
  the composite spectrum at $\lambda_r < 912$\AA\ having
  $\mlmfp = 150\mhmpc$ but would greatly overpredict \tlya\ and \tlyman.
}
\label{fig:wfc3}
\end{figure}

Figure~\ref{fig:wfc3} compares the $z = 2.44$
composite spectrum of O13 with the `best' model having $\mlmfp = 152
\mhmpc$. We find that the most extreme tilt\footnote{We consider even
  smaller \dat\ values in the next section and note here that these
  greatly overpredict the Lyman series opacities.} of the quasar SED ($\mdat = -0.8$)
is favored, but even with this rather extreme SED the resultant model
is a very poor description of the data.  If we adopt the RMS of the
composite flux assessed from a bootstrap analysis (O13; their figure 8) 
and assume a Gaussian PDF, 
we measure $\chi_{\nu, \rm min}^2 = 2.22$ for the $\nu = 33$ degrees of
freedom from the 37 pixels spanning $\lambda_r = 650-910$\AA.
This implies that the model is ruled out at high significance
(99.99\%c.l).
Note however that
the scatter in the composite is not truly Gaussian and
the flux is highly correlated by the nature of continuum opacity.
Nevertheless, we conclude that the \lmfp\ value inferred by R13 either
violates the O13 quasar composite spectrum or requires a tilt to the
SED that contradicts other measurements (see also
$\S$~\ref{sec:sys_error}).

\begin{table*}
\centering
\begin{minipage}{170mm}
\caption{\fnhi\ Constraints at $z \approx 2.5$}
\label{tab:fn_constraints}
\begin{tabular}{ccccll}
\hline
Constraint &$z^a$ & log \nhi & Value$^b$ & Comment
& Reference \\
\hline
Lya Forest&2.30&12.75--13.00&$-11.08^{+  0.02}_{-  0.01}$&Grabbed from astro-ph on August 30, 2013&K13\\
&&13.00--13.25&$-11.44^{+  0.02}_{-  0.02}$\\
&&13.25--13.50&$-11.80^{+  0.02}_{-  0.02}$\\
&&13.50--13.75&$-12.16^{+  0.02}_{-  0.02}$\\
&&13.75--14.00&$-12.51^{+  0.03}_{-  0.02}$\\
&&14.00--14.25&$-12.99^{+  0.03}_{-  0.03}$\\
&&14.25--14.50&$-13.46^{+  0.05}_{-  0.04}$\\
&&14.50--14.75&$-13.78^{+  0.05}_{-  0.04}$\\
&&14.75--15.06&$-14.29^{+  0.07}_{-  0.06}$\\
&&15.06--15.50&$-14.93^{+  0.06}_{-  0.05}$\\
&&15.50--16.00&$-15.97^{+  0.12}_{-  0.10}$\\
&&16.00--16.50&$-16.67^{+  0.17}_{-  0.12}$\\
&&16.50--17.00&$-17.22^{+  0.18}_{-  0.12}$\\
&&17.00--17.50&$-17.90^{+  0.23}_{-  0.15}$\\
&&17.50--18.00&$-19.17^{+  0.40}_{-  0.30}$\\
Lya Forest&2.37&13.01--13.11&$-11.33^{+  0.02}_{-  0.03}$&Provided by G. Rudie on Aug 28, 2013&R13\\
&&13.11--13.21&$-11.48^{+  0.02}_{-  0.03}$\\
&&13.21--13.32&$-11.65^{+  0.02}_{-  0.03}$\\
&&13.32--13.45&$-11.76^{+  0.02}_{-  0.02}$\\
&&13.45--13.59&$-11.96^{+  0.02}_{-  0.03}$\\
&&13.59--13.74&$-12.23^{+  0.03}_{-  0.03}$\\
&&13.74--13.90&$-12.53^{+  0.03}_{-  0.03}$\\
&&13.90--14.08&$-12.78^{+  0.03}_{-  0.03}$\\
&&14.08--14.28&$-13.11^{+  0.03}_{-  0.04}$\\
&&14.28--14.50&$-13.41^{+  0.04}_{-  0.04}$\\
&&14.50--14.73&$-13.74^{+  0.04}_{-  0.04}$\\
&&14.73--15.00&$-14.20^{+  0.05}_{-  0.05}$\\
&&15.00--15.28&$-14.75^{+  0.06}_{-  0.07}$\\
&&15.28--15.60&$-15.19^{+  0.07}_{-  0.08}$\\
&&15.60--15.94&$-15.66^{+  0.08}_{-  0.09}$\\
&&15.94--16.32&$-16.25^{+  0.09}_{-  0.12}$\\
&&16.32--16.73&$-16.91^{+  0.11}_{-  0.16}$\\
&&16.73--17.20&$-17.37^{+  0.11}_{-  0.15}$\\
Lya Forest&2.34&12.50--13.00&$-11.16^{+  0.04}_{-  0.04}$&Recalculated for our Cosmology                       &K02  \\
&&13.00--13.50&$-11.73^{+  0.05}_{-  0.05}$\\
&&13.50--14.00&$-12.51^{+  0.07}_{-  0.07}$\\
&&14.00--14.50&$-13.27^{+  0.10}_{-  0.09}$\\
SLLS      &2.51&19.00--19.60&$-20.60^{+  0.19}_{-  0.17}$&Only 30 systems total                                &OPB07\\
&&19.60--20.30&$-21.48^{+  0.23}_{-  0.20}$\\
DLA       &2.51&20.30--20.50&$-21.80^{+  0.06}_{-  0.06}$&$z=[2.3,2.7]$; modest SDSS bias \citep[see][]{np+09}?&PW09 \\
&&20.50--20.70&$-22.26^{+  0.08}_{-  0.08}$\\
&&20.70--20.90&$-22.38^{+  0.08}_{-  0.07}$\\
&&20.90--21.10&$-22.85^{+  0.11}_{-  0.10}$\\
&&21.10--21.30&$-23.33^{+  0.15}_{-  0.15}$\\
&&21.30--21.50&$-23.58^{+  0.16}_{-  0.15}$\\
&&21.50--21.70&$-24.30^{+  0.34}_{-  0.30}$\\
&&21.70--21.90&$-24.98^{+  0.76}_{-  0.52}$\\
&&21.90--22.10&$-99.00^{+-99.00}_{--24.60}$\\
&&22.10--22.30&$-99.00^{+-99.00}_{--24.80}$\\
\hline
\end{tabular}
\end{minipage}
\end{table*}
\begin{table*}
\begin{minipage}{170mm}
\contcaption{\fnhi\ Constraints at $z \approx 2.5$}
\centering
\begin{tabular}{ccccll}
DLA&2.50&20.10--20.20&$-21.49^{+  0.05}_{-  0.05}$&Tablulated&N12\\
&&20.20--20.30&$-21.61^{+  0.05}_{-  0.05}$\\
&&20.30--20.40&$-21.70^{+  0.05}_{-  0.05}$\\
&&20.40--20.50&$-21.84^{+  0.05}_{-  0.05}$\\
&&20.50--20.60&$-22.00^{+  0.05}_{-  0.05}$\\
&&20.60--20.70&$-22.16^{+  0.05}_{-  0.05}$\\
&&20.70--20.80&$-22.34^{+  0.05}_{-  0.05}$\\
&&20.80--20.90&$-22.53^{+  0.05}_{-  0.05}$\\
&&20.90--21.00&$-22.69^{+  0.05}_{-  0.05}$\\
&&21.00--21.10&$-22.93^{+  0.05}_{-  0.05}$\\
&&21.10--21.20&$-23.13^{+  0.05}_{-  0.05}$\\
&&21.20--21.30&$-23.30^{+  0.05}_{-  0.05}$\\
&&21.30--21.40&$-23.60^{+  0.06}_{-  0.06}$\\
&&21.40--21.50&$-23.83^{+  0.07}_{-  0.07}$\\
&&21.50--21.60&$-24.03^{+  0.08}_{-  0.08}$\\
&&21.60--21.70&$-24.22^{+  0.08}_{-  0.08}$\\
&&21.70--21.80&$-24.64^{+  0.12}_{-  0.12}$\\
&&21.80--21.90&$-24.87^{+  0.18}_{-  0.18}$\\
&&21.90--22.00&$-25.62^{+  0.53}_{-  0.53}$\\
&&22.00--22.20&$-26.07^{+  0.53}_{-  0.53}$\\
&&22.20--22.40&$-26.27^{+  0.53}_{-  0.53}$\\
\tlox&2.23&$> 17.49$&$  0.30\pm  0.07$& &OPW12\\
\tlya&2.40&12.00--17.00&$ 0.198\pm 0.007$&Converted to \tlya\ from $D_A$.  No LLS, no metals.&K05\\
\lmfp&2.44&12 -- 22&$ 242\pm 41\, \mhmpc $& &OPW13\\
\hline
\end{tabular}
\end{minipage}
{$^a$}Effective redshift where the constraint was determined. \\
{$^b$}\fnhi\ constraints are given in log. \\
Note: there were two modifications to the \fnhi\ data from N12 (see text).\\
KT97: \cite{kt97}; K01: \cite{kim+01}; K02: \cite{kim02}; K05: \cite{kts+05}; OPB07: \cite{opb+07}; PW09: \cite{pw09}; OPW13: \cite{omeara13}; N12: \cite{noterdaeme+12}
\end{table*}

\subsection{Tension in \fnhi}
\label{sec:fnhi}

Ostensibly,
the above analysis suggests that the R13 and O13 data lie in
strong conflict.  
Thus far, however, we have only considered the \lmfp\ and
$\gell$ values reported/assumed by R13 and not their
actual measurements or model of \fnhi.  
In this spirit, we
perform a joint analysis of \fnhi\ imposing all of the constraints
included by O13 and adding the R13 measurements. 
As a reminder,
the constraints adopted by O13 were
(see also Table~\ref{tab:fn_constraints}):
 (1) the O13 estimate of \lmfp\ revised for cosmology
 (Figure~\ref{fig:compare_mfp});
 (2) the mean opacity of the \lya\ forest \citep{kts+05};
 (3) the incidence of $\tau \ge 2$ LLS \citep[][O13]{ribaudo11};
 (4) the \fnhi\ measurements of \citet[][hereafter K02]{kim02};
 (5) the \fnhi\ measurements of strong absorption systems from 
 \citet{opb+07} and \citet{pw09}.

Before proceeding to model the combined R13 and O13 constraints,
we examine the \fnhi\ models published by R13 and O13 tested by one
another's data.  
Each model, of course, gives good statistical results ($\chi_\nu^2
\approx 1$) for fits to the data analyzed in each paper.
R13 favored a 4-parameter, disjoint set of two power-laws
split at $\mnhi = 10^{15.14} \cm{-2}$.  This model yields an
acceptable
$\chi^2_\nu = 1.44$ for their own measurements, but gives $\chi^2_\nu = 22.3$ for
the O13 constraints alone \citep[including][]{kim02} and
$\chi^2_\nu = 11.3$ for the combined constraints (O13 plus R13). 
Here a substantial contribution to the $\chi^2$ is from the absorption 
systems with largest \nhi\ values, which R13 did not include in their
analysis. Therefore, the R13 model is ruled out at very high confidence.  
Similarly, the O13 model (a 6-parameter, continuous set of power-laws)
gives $\chi^2_\nu = 11.1$ for the combined datasets
driven entirely by the R13 measurements
(especially those at low \nhi\ values).  

\begin{figure}
\includegraphics[width=3.7in]{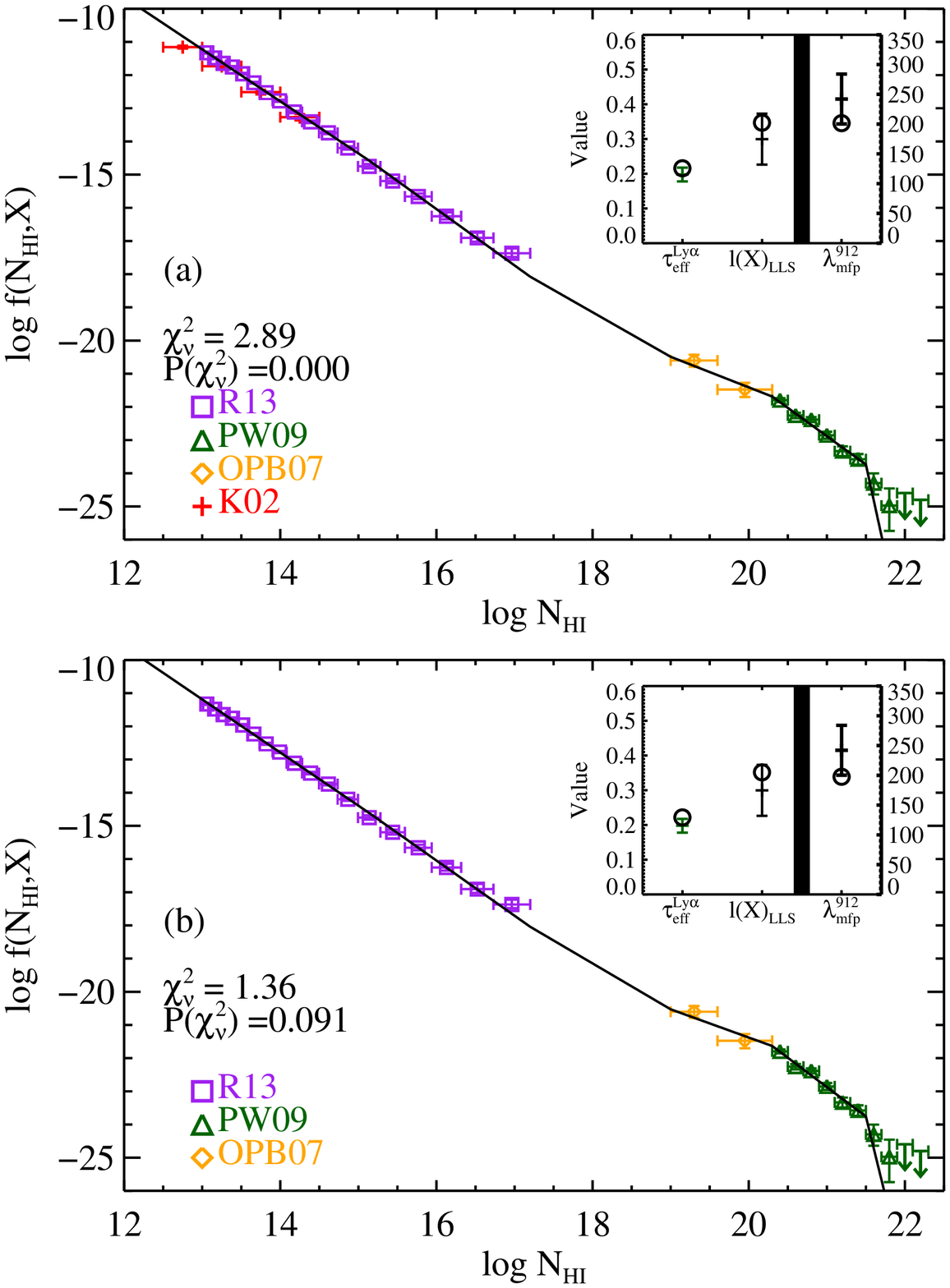}
\caption{(a) The black curve shows a 7-parameter continuous, power-law
  fit with MCMC techniques to the \fnhi\ constraints labeled
  \citep[Table~\ref{tab:fn_constraints}][]{rudie13,pw09,opb+07,kim02} 
  and the integral constraints in the inset \citep{kts+05,omeara13}.
  This model yields an unacceptably high reduced $\chi^2_\nu$, driven
  primarily by the \fnhi\ values at low \nhi.  
  (b) Same as (a) but without the K02 measurements.  The resultant
  $\chi_\nu^2$ may be considered acceptable, but we stress that this
  model is under great tension between the \fnhi\ measurements at
  $\mnhi \approx 10^{17} \cm{-2}$, the $\ell(X)$ constraint, and the
  \lmfp\ value.  
}
\label{fig:failed_fn}
\end{figure}

Given these `failed' models, one is motivated to ask whether any 
\fnhi\ model can fit all of the available data at $z \approx 2.5$.
We proceed to fit the constraints used by O13 together with the
measurements of R13 with a continuous series of power-laws with
log normalization $k_{12}$ at $\mnhi = 10^{12} \cm{-2}$ and with 
breaks at five \nhi\ `pivots' motivated by the data:
$\log \mnhi^{\rm pivot} = [15.14, 17.2, 19.0, 20.3, 21.5]$.
Each segment has a slope $\beta$ labeled by the pivot.

We fit this 7-parameter model\footnote{The model assumes a redshift
  evolution $f(\mnhi,X) \propto (1+z)^{\gell}$ with $\gell = 1.5$ for
  all \nhi.  We considered models with $\gell$ as a free parameter, but
  the observations considered offer very little constraint.}
to the observations using a Markov Chain Monte Carlo (MCMC;
Metropolis-Hastings) algorithm, employing a \vstep\ step-size for each
parameter.  Random initializations and multiple long MCMC chains were
generated to insure proper convergence.  Figure~\ref{fig:failed_fn}a presents the
best-fit model (see Table~\ref{tab:fn_results})
which has a
$\chi^2_\nu=2.89$ with a probability $P(\chi_\nu^2)< 10^{-5}$
indicating a very poor description of the observations.  The
deviation is driven primarily by the \fnhi\ values of K02 at low \nhi.
Therefore, we repeat the analysis without the K02 measurements,
noting that the R13 data cover the lower \nhi\ range on their own,
and recover the model shown in Figure~\ref{fig:failed_fn}b.
The $\chi^2_\nu$ is notably improved and may even be considered
acceptable, but the figure also emphasizes the tension between R13 and
O13.  The \lmfp, $\ell(X)_{\tau \ge 2}$, and $\mfnhi \gtrsim 10^{16.5}
\cm{-2}$ measurements are all poorly described by this model\footnote{
  We considered one further model of these data -- an 8 parameter
  power-law with an additional pivot at $\mnhi^{\rm pivot} = 10^{18}
  \cm{-2}$ -- which yields a satisfactory \lmfp\ value but predicts a
  very shallow slope $\beta_{18} > -0.4$ at $\mnhi \approx 10^{18.5}
  \cm{-2}$ which we disfavor.}. 
We conclude that there is substantial tension between the various
measurements of the IGM when one attempts to model these with a single
\fnhi\ model.

\begin{table}
\caption{\fnhi\ MCMC Results \label{tab:fn_results}}
\begin{tabular}{ccccc}
\hline
Parameter &Prior & Median & 16th\% & 84th\% \\
\hline
\multicolumn{5}{c}{Results for R13 and O13 (Figure~\ref{fig:failed_fn}a) } \\
$k_{12}$&$U(-\infty,\infty)$&$ -9.65$& 0.03& 0.02\\
$\beta_{12}$&$U(-\infty,\infty)$&$ -1.57$& 0.01& 0.02\\
$\beta_{15.14}$&$U(-\infty,\infty)$&$ -1.68$& 0.05& 0.05\\
$\beta_{17.2}$&$U(-\infty,\infty)$&$ -1.39$& 0.12& 0.09\\
$\beta_{19}$&$U(-\infty,\infty)$&$ -0.83$& 0.12& 0.13\\
$\beta_{20.3}$&$U(-\infty,\infty)$&$ -1.75$& 0.14& 0.09\\
$\beta_{21.5}$&$U(-\infty,\infty)$&$-11.65$& 1.64& 6.35\\
\multicolumn{5}{c}{Results for R13 and O13 without K02 (Figure~\ref{fig:failed_fn}b) } \\
$k_{12}$&$U(-\infty,\infty)$&$ -9.59$& 0.02& 0.03\\
$\beta_{12}$&$U(-\infty,\infty)$&$ -1.60$& 0.02& 0.01\\
$\beta_{15.14}$&$U(-\infty,\infty)$&$ -1.66$& 0.04& 0.05\\
$\beta_{17.2}$&$U(-\infty,\infty)$&$ -1.42$& 0.11& 0.11\\
$\beta_{19}$&$U(-\infty,\infty)$&$ -0.82$& 0.12& 0.13\\
$\beta_{20.3}$&$U(-\infty,\infty)$&$ -1.74$& 0.14& 0.09\\
$\beta_{21.5}$&$U(-\infty,\infty)$&$-12.86$& 2.71& 7.45\\
\multicolumn{5}{c}{Results for 8-parameter model (R13 and O13 without K02) } \\
$k_{12}$&$U(-\infty,\infty)$&$ -9.58$& 0.03& 0.03\\
$\beta_{12}$&$U(-\infty,\infty)$&$ -1.61$& 0.02& 0.01\\
$\beta_{15.14}$&$U(-\infty,\infty)$&$ -1.64$& 0.02& 0.08\\
$\beta_{17.2}$&$U(-\infty,\infty)$&$ -2.11$& 1.30& 0.33\\
$\beta_{18.0}$&$U(-\infty,\infty)$&$ -0.50$& 0.56& 1.29\\
$\beta_{19.0}$&$U(-\infty,\infty)$&$ -1.04$& 0.20& 0.19\\
$\beta_{20.3}$&$U(-\infty,\infty)$&$ -1.80$& 0.07& 0.18\\
$\beta_{21.5}$&$U(-\infty,\infty)$&$ -9.64$& 1.18& 1.97\\
\multicolumn{5}{c}{Results for Spline Model (Figure~\ref{fig:final_fn}) } \\
$k_{12}$&$U(-\infty,\infty)$&$ -9.72$& 0.04& 0.07\\
$k_{15}$&$U(-\infty,\infty)$&$-14.41$& 0.02& 0.02\\
$k_{17}$&$U(-\infty,\infty)$&$-17.94$& 0.08& 0.13\\
$k_{18}$&$U(-\infty,\infty)$&$-19.39$& 0.16& 0.10\\
$k_{20}$&$U(-\infty,\infty)$&$-21.28$& 0.03& 0.03\\
$k_{21}$&$U(-\infty,\infty)$&$-22.82$& 0.02& 0.02\\
$k_{21.5}$&$U(-\infty,\infty)$&$-23.95$& 0.03& 0.04\\
$k_{22}$&$U(-\infty,\infty)$&$-25.50$& 0.15& 0.10\\
\hline
\end{tabular}
\end{table}

\begin{figure}
\includegraphics[width=3.2in,angle=90]{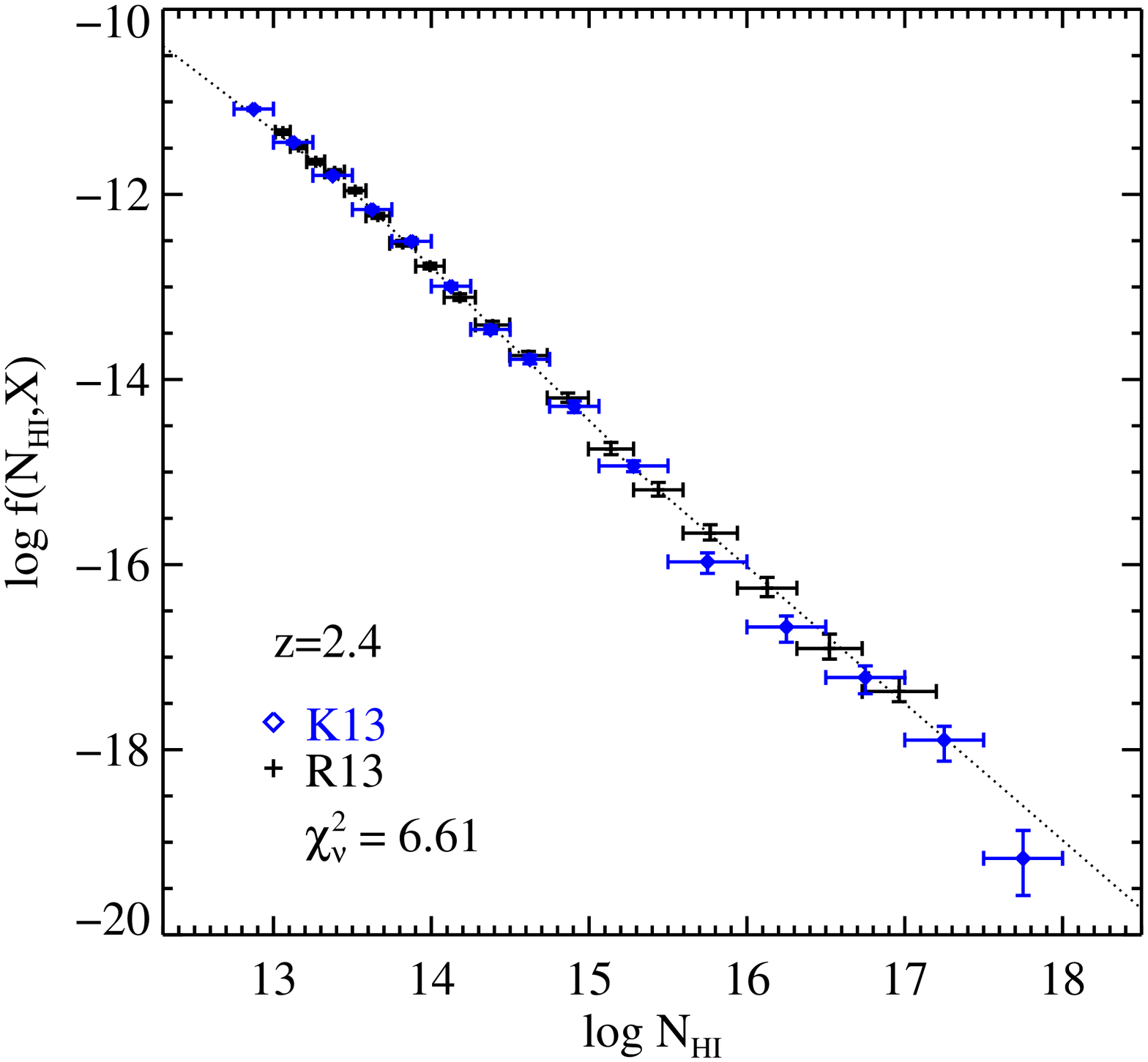}
\caption{Comparison of the independent \fnhi\ measurements for the IGM
  at $z=2.4$ by 
  \citet[][K13; blue]{kim13} and \citet[][R13; black]{rudie13}.  Although the two
  groups use similar techniques on similar quality spectra (with a few
  notable differences), the two sets of results are highly
  inconsistent with one another, especially at the extrema.  For
  example, the best-fit \fnhi\ model of R13 (dotted line) has a
  $\chi_\nu^2 = 6.61$ and $P(\chi_\nu^2) < 10^{-5}$ when applied to
  the K13 measurements.  
  This model also fails when restricting the comparison to $\mnhi >
  10^{15.5} \cm{-2}$.
  We conclude that there is a significant,
  systematic uncertainty in assessing \fnhi\ at $\mnhi > 10^{15}
  \cm{-2}$ that was unaccounted for by these authors. 
}
\label{fig:r13_k13}
\end{figure}

In fact,
the situation becomes untenable if one includes the recent
measurements
of \fnhi\ published by \citet[][hereafter K13]{kim13}.
Those authors performed a similar line-profile fitting analysis to R13
of high S/N, echelle quasar spectra using standard Voigt-profile
fitting techniques.  They considered fewer constraints
from higher order Lyman series lines than R13, but argued that this
had minimal effect on their results.  
Figure~\ref{fig:r13_k13} compares the two datasets.  The values are in
reasonably good agreement at modest \nhi\ values when \lya\ and/or
\lyb\ are unsaturated.  At larger and lower \nhi\ values, however, the
two sets of measurements are highly inconsistent.
For example, the \fnhi\ model of R13 yields a $\chi_\nu^2$ on
the K13 measurements which implies the two measurements disagree
at greater than $99.999\%$c.l. (Figure~\ref{fig:r13_k13}). 
Even if we restrict the comparison to measurements with $\mnhi >
10^{15.5} \cm{-2}$, the R13 model gives
$\chi_\nu^2 = 3$ and $P(\chi^2_\nu) = 0.018$.  
Not surprisingly, we cannot find any \fnhi\ model that satisfactorily
fits the suite of K13, R13, and O13 constraints on the IGM at $z
\approx 2.5$.

\section{Resolutions}
\label{sec:resolve}

We explore three ways to reconcile the conflict among the
observational constraints of the $z \approx 2.5$ IGM, as described in the previous
section.  Each of these may contribute to a resolution:
 (1) O13 have overestimated \lmfp\ by underestimating the spectral
 slope of the average quasar SED and/or by suffering from a statistical
 fluctuation;
 (2) line-blending has led to the double counting of absorption
 systems and also implies a substantial systematic uncertainty for \fnhi;
 (3) the clustering of absorption systems with $\mtll \gtrsim 1$ drives the evaluation of 
 Equation~\ref{eqn:teff} to underestimate \lmfp.


\begin{figure}
\includegraphics[width=3.1in,angle=90]{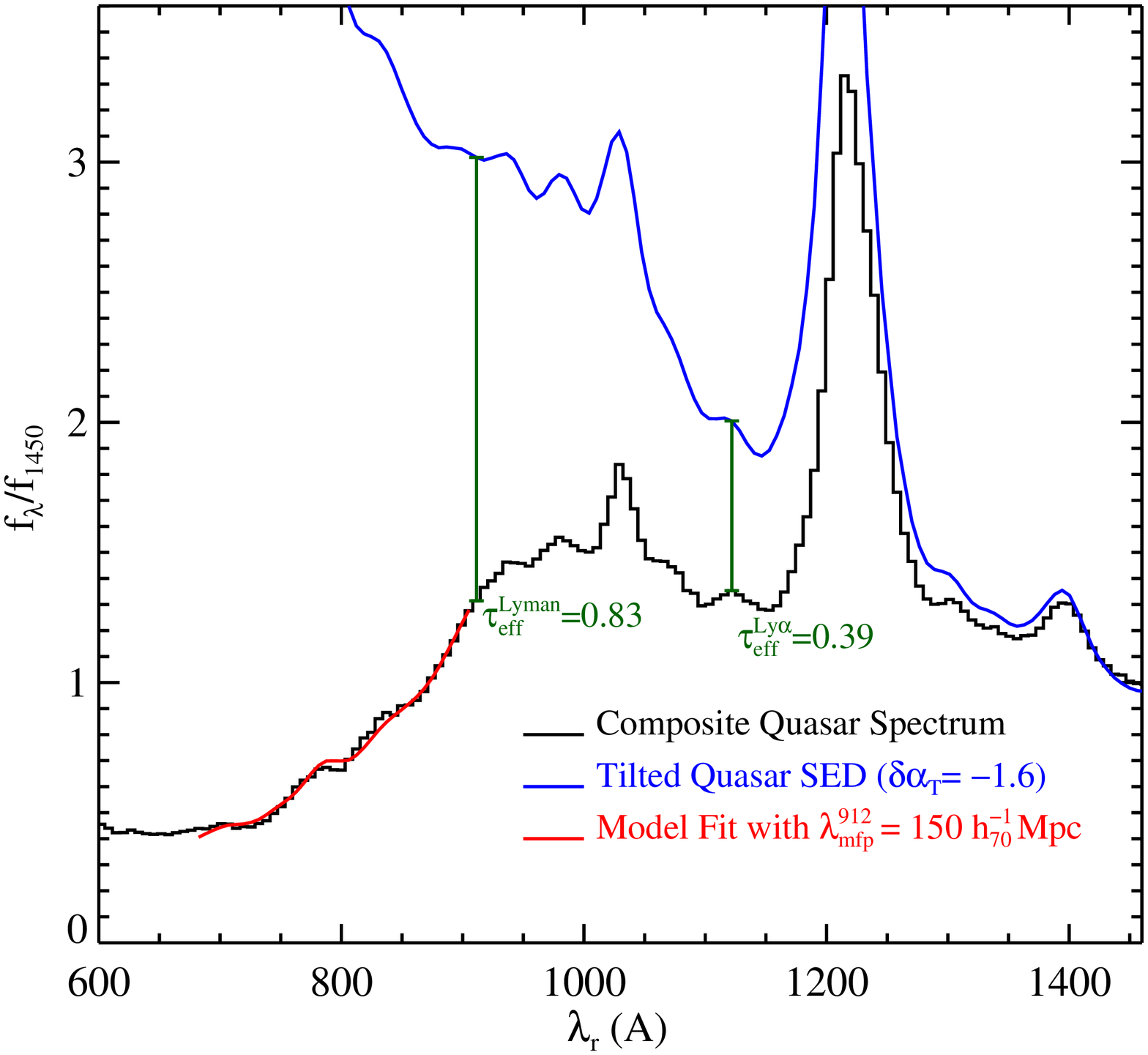}
\caption{Similar to Figure~\ref{fig:wfc3} but allowing the quasar
  SED tilt \dat\ to hold any value.  In this case, the quasar
  composite spectrum (black) may be modeled by an IGM with $\mlmfp =
  150 \mhmpc$ (red curve) provided $\mdat < -1.5$.  
  However, the resultant values for the
  effective optical depths of \lya\ and the full Lyman series (\tlya,
  \tlyman) vastly exceed previous measurements \citep{kts+05,omeara13}.  
  Furthermore, the implied
  quasar spectral slope $f^{\rm QSO}_\nu \propto \nu^{0.1}$ is steeper
  than any previous estimation and over-predicts the X-ray to optical
  ratio by over an order-of-magnitude.
}
\label{fig:tilt}
\end{figure}

\subsection{Additional Error}
\label{sec:sys_error}

Regarding the O13 analysis,
it is possible to achieve a good model of the composite spectrum with
$\mlmfp = 150 \, \mhmpc$ if one allows for an even more extreme
tilt\footnote{ 
  We also note that O13 underestimated $\gamma_\tau$, but find that a
  larger value actually favors a slightly larger \lmfp\ value.}
of the intrinsic quasar SED, i.e.\ $\mdat < -1.5$
(Figure~\ref{fig:tilt}).
This same model, however, overpredicts both the
average \lya\ opacity \tlya\ at $\lambda_r \approx 1120$\AA\
and especially
the integrated Lyman series opacity \tlyman\ at 
$\lambda_r \approx 912$\AA. 
Furthermore, this quasar SED would have $f_\nu \propto \nu^\alpha$
with $\alpha > 0$ in the far-UV,
exceeding any plausible estimation for quasars \citep{lusso13}.
Therefore, we strongly disfavor this explanation.


Systematic error in both the K13 and R13 studies is a major concern, especially in
light of the large differences between the two measurements 
(Figure~\ref{fig:r13_k13}). As discussed in $\S$~\ref{sec:fnhi}, 
there is likely a large systematic error in evaluating
\fnhi\ at $\mnhi \approx 10^{16} \cm{-2}$ with
traditional line-fitting.  
This is evident simply from the dispersion in results from the various
studies. 
We conclude that there is substantial systematic error in assessing
lines on the flat portion of the curve-of-growth which was
not accounted for by
these authors. 
It could be related to line-saturation (i.e.\ limited coverage of the
complete Lyman series), line-blending (see below), 
or even to sample variance.
Blind analysis of mock spectra may help to
resolve these issues and we encourage such a study. We also recommend
that future works
restrict their analysis to absorption systems with spectra that cover
down to at least Ly$\epsilon$ (see the Appendix of R13), 
instead of only \lya\ and \lyb\ as done
previously.   Finally, a dataset of $\sim 100$ sightlines may be
required to properly account for sample variance.

Allowing for a larger uncertainty in the \fnhi\ measurements at $\mnhi
\gtrsim 10^{15} \cm{-2}$, it may be possible to find a model which
describes well the suite of IGM constraints
(Table~\ref{tab:fn_constraints}).  
Consider the following simple approach.
We measure the average offset
between the R13 model of $\log \mfnhi$ and the K13 measurements at $\mnhi \ge
10^{15.5} \cm{-2}$ to be 0.25\,dex.
We can then use the average of the K13 measurements and the
R13 model\footnote{We use the R13 measurements at $\mnhi <
  10^{15.5} \cm{-2}$.} 
at $\mnhi > 10^{15.5} \cm{-2}$ and impose an additional
0.25\,dex uncertainty, added in quadrature.  
We have also adopted the \citet[][hereafter N12]{noterdaeme+12}
measurements of \fnhi\ for absorption systems with $\mnhi \ge 10^{20}
\cm{-2}$ with two modifications (approved by the lead authors of N12):
(i) we ignore their first data point (at $\mnhi = 10^{20} \cm{-2}$)
which is likely biased by incompleteness;
(ii) we adopt a minimum uncertainty of 0.05\,dex to account for
systematics.  
Lastly, we introduce a new functional form for \fnhi, monotonically
declining spline using the cubic Hermite spline algorithm of
\cite{fritsch80}. 
We parameterize this spline with 8 points which are only allowed to
vary in amplitude (see Table~\ref{tab:fn_results}).
With all of these modifications,
we recover an \fnhi\ model that 
reasonably describes all of the data.  
Despite this model's success, we
now argue that the traditional \fnhi\ formalism is invalid in light of
the clustering of absorption-line systems.

\begin{figure}
\includegraphics[width=3.3in,angle=90]{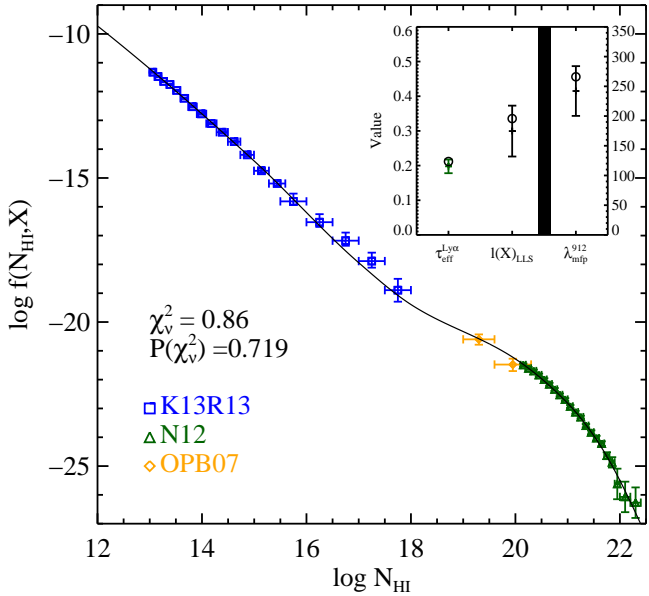}
\caption{The black curve shows the best-fit model of \fnhi\ to all of
  the constraints presented (see Table~\ref{tab:fn_constraints}).
  For the \fnhi\ measurements, we have adopted the values of R13 at
  $\mnhi < 10^{15.5} \cm{-2}$ but have combined their model with the
  K13 values at higher \nhi\ and adopted a larger uncertainty (see
  text).  The resultant model is a reasonable description of the data.
}
\label{fig:final_fn}
\end{figure}

\subsection{Line-blending and Absorption System Clustering}

Although the tension in IGM measurements may be largely explained by the
above statistical and systematic uncertainties, we believe that the
third effect (clustering) plays as great a role in explaining the
apparent discrepancies.  Recently,
\cite{qpq6} have measured a remarkably large clustering
amplitude between quasars \citep[which reside in massive
halos;][]{white12} and LLS: $r_0^{\rm LLS} > 10\, h^{-1}_{100} \rm \,
Mpc$ \citep[see also][]{qpq2}. They further argued that a significant fraction 
of LLS (possibly all!) occur within one proper Mpc of massive and
(i.e.\ rare) dark matter halos.  This implies
that optically thick gas is not randomly distributed throughout the
IGM, but instead occupies a smaller portion of the volume.  
It also implies that one will more frequently 
discover multiple, strong absorption systems at small velocity
separations.

The clustering of absorption systems impacts the relation between
\fnhi\ and mean transmission of the IGM. 
First, the clustering of absorption systems leads to ``line-blending''
-- cases where
two or more absorption systems occur within a small velocity separation.  When this
occurs for lines having $\mnhi \gtrsim 10^{16.5} \cm{-2}$, the
combined system has $\mtll \gtrsim 1$, and a survey of 
Lyman continuum opacity would also classify it as an LLS.  
This leads to {\it the double counting of Lyman limit
opacity} when combining multiple studies at varying resolution, 
and a standard analysis (i.e.\ Equation~\ref{eqn:teff})
would overestimate \teff\ and underestimate \lmfp.  

\begin{figure}
\includegraphics[width=3.5in]{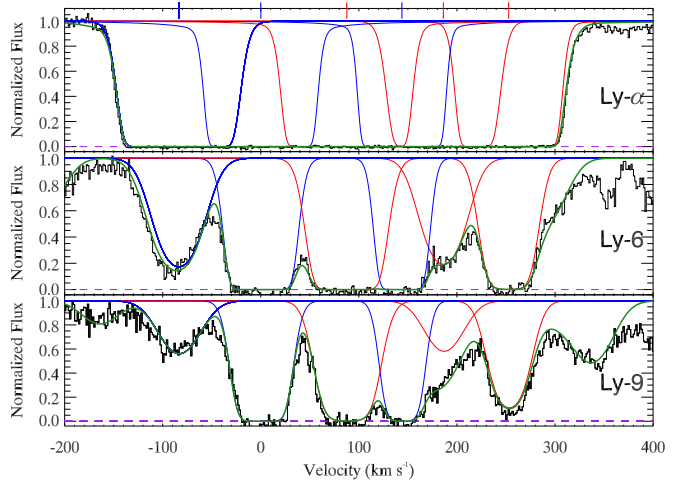}
\caption{Lyman series absorption for the $z \sim 2.45$ LLS towards
  Q1549+1919.  The three absorbers shown in red have their \nhi\
  fixed to the values presented in the appendix of R13 for this LLS.
  R13 do not provide exact redshifts or velocity widths for these
  absorbers, so they have been chosen to best match by eye the
  absorption presented in R13 while still presenting a good overall
  absorption model to the LLS. In addition, we show in blue all
  additional absorbers with $\log \mnhi>15.0$ that are needed to fit the
  strongest absorption features in the LLS.  The green curve shows the
  full absorption model, which includes additional $\log \mnhi <15.0$
  absorbers to provide a good fit to the data.  In total, six
  absorbers, including three with $\log\mnhi>16.5$.
}
\label{fig:model}
\end{figure}

The concept of line-blending is nicely illustrated by the R13 data.
For example. consider the absorption complex at $z\sim2.45$ toward
Q1549+1919.
Analysis of the full Lyman series reveals a complex 
system\footnote{R13 present three absorbers from this LLS with $\mnhi
  =10^{15.79}, 10^{16.32}$, and $10^{16.82}\cm{-2}$, but do not provide
  redshifts or velocity widths.  We use these column densities as part
  of the model presented in Figure~\ref{fig:model}.   As R13 did not publish
  their linelists, we cannot confirm if they included additional
  absorbers in this LLS or any other LLS in their sample with multiple
  strong components, in their sample for \fnhi.
}
of six absorbers having $\mnhi >10^{15} \cm{-2}$, all within $\Delta
v=400 \mkms$.  These lines have individual column densities, in
increasing strength, of $\mnhi =10^{15.79}, 10^{15.85}, 10^{16.32},
10^{16.70}, 10^{16.82}$ and $10^{17.00}\cm{-2}$.  Although there are
significant degeneracies between components in the model, one thing is
certain:  multiple strong ($\mnhi > 10^{15.5} \cm{-2}$) \ion{H}{I} absorbers are
required to account for the observed absorption (Figure~\ref{fig:model}).  If
assessed independently via Lyman continuum opacity, the complex would
be recorded as an LLS with $\mnhi =10^{17.4} \cm{-2}$ and would
additionally contribute to \fnhi\ at this higher \nhi\ value.
We conclude that R13 underestimated \lmfp\ because of the double
counting of LL opacity; indeed, this is also true of all previous
authors that coupled \fnhi\ and LLS statistics.
We note further that 
  line-blending may have also impacted the \fnhi\ model of
  O13, who reported a deficit of $\mnhi \approx 10^{17} \cm{-2}$ systems
  relative to R13.  The O13 conclusion was based on their \lmfp\ value
  and the incidence of LLS.  
  As such, clusters of $\mnhi = 10^{16} \cm{-2}$ lines were included
  as LLS and fewer such systems
  were required to match the \lmfp\ measurement.

Second (and similarly), the clustering of absorption systems means
optically thick gas is not randomly distributed throughout the universe.
This contradicts the standard formalism (Equation~\ref{eqn:teff})
used to calculate \teff\
which assumes a Poisson distribution in the IGM. For accurate results,
one must fully account for clustering to use \fnhi\ as a
description of the IGM.
A full and proper treatment must await
future observations, in tandem with the analysis of cosmological
simulations that include hydrodynamics and radiative transfer. 
For now, we offer below some insight
on how the clustering of LLS will tend to increase the 
\lmfp.

\subsection{Modifying the Opacity of the IGM for the Clustering of
  LLS}
\label{sec:toy}

Equation~\ref{eqn:teff} for the effective continuum optical depth of a
clumpy IGM is valid under the assumption of a random distribution of
absorbers along the line of sight. The formula can be easily
understood if we consider a situation in Euclidean space in which all
absorbers have the same optical depth $\tau_0$, and the mean number of
systems along the path is $\bar N$. In this case the Poissonian
probability of encountering a total optical depth $N\tau_0$ along the
path (with $N$ integer) is $p(N)$, where  
\begin{equation}
p(N)=e^{-\bar N} {{\bar N}^N \over N!}. 
\end{equation}
The mean attenuation is then  
\begin{eqnarray}
\langle e^{-\tau}\rangle & =&\sum\limits_{N=0}^\infty e^{-N\tau_0}p(N)=
e^{-\bar N} \sum\limits_{N=0}^\infty {(e^{-\tau_0} {\bar N})^N\over N!} \nonumber \\
& =&e^{-{\bar N} (1-e^{-\tau_0})},
\end{eqnarray}
and the effective optical depth is $\tau_{\rm eff}=-\ln \langle e^{-\tau}\rangle={\bar N} (1-e^{-\tau_0})$ (cf. Equation 1). 
When $\tau_0\ll 1$, $\tau_{\rm eff}$ becomes equal to the mean optical depth. In the opposite limit, the obscuration is picket-fence, and
the effective optical depth becomes equal to the mean number of optically thick absorbers along the line of sight. 

\begin{figure}
\centering
\includegraphics[width=0.59\textwidth]{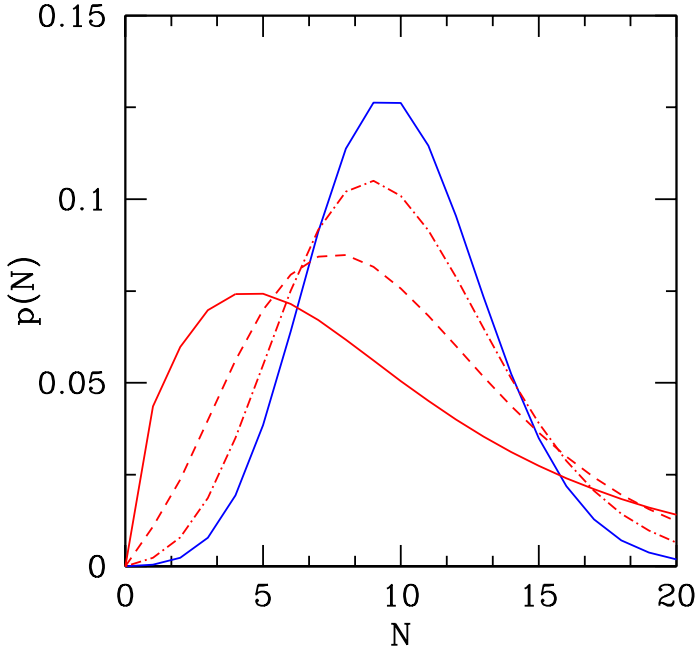}
\vspace{0.0cm}
\caption{The frequency distribution $p(N)$ for $\bar N=10$ and $N=0, 1, 2, 3,...$. The Poisson limit is shown with the blue solid curve, 
while the highly clustered distributions for $b=0.2, 0.4, 0.6$ are shown with the red dot-dashed, dashed, and solid curves, respectively. 
}
\label{fig1}
\end{figure}

In the evaluation of Equation~\ref{eqn:teff}, then lines cluster
together it is the total \nhi\ that contributes.
To assess the impact of gravitational clustering on the effective opacity and therefore on the mean free path 
of ionizing radiation through the IGM, let us assume instead a probability distribution function of the form
\begin{equation}
p(N)={{\bar N}(1-b)\over N!}[{\bar N}(1-b)+Nb]^{N-1}e^{-{\bar N}(1-b)-Nb}.
\label{eq:nP}
\end{equation}
Predicted by gravitational thermodynamics to describe galaxy clustering in an expanding universe 
\citep{Saslaw84,Saslaw89}, this function reduces to a 
Poisson distribution (no gravitational interactions) when the parameter $b=-W/2K$, which measures the degree of virialization, 
is $b=0$. Figure \ref{fig1} depicts the frequency distribution $p(N)$ for $\bar{N}=10$ and $b=0$ (Poisson), 0.2, 0.4, and 0.6.
The extreme non-Poisson limit corresponds to $b\rightarrow 1$, while the first moment of the distribution, 
$\langle (\Delta N)^2\rangle^{1/2}\equiv \langle (N-{\bar N})^2\rangle^{1/2}=\sqrt{N}/(1-b)$ shows that correlated 
fluctuations are amplified over the Poisson value by the factor $(1-b)^{-1}$.

\begin{figure}
\centering
\includegraphics[width=0.59\textwidth]{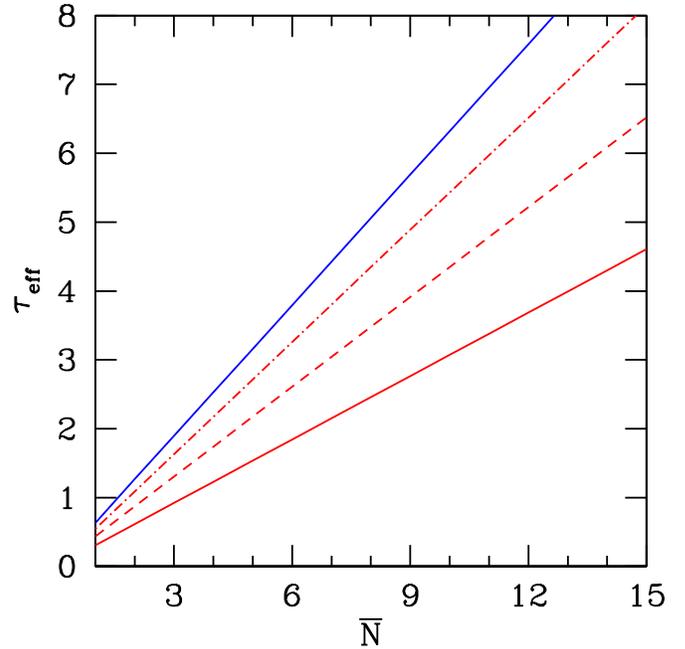}
\vspace{0.0cm}
\caption{The effective opacity in Euclidean space corresponding to the frequency distribution of equation (\ref{eq:nP}), as a function 
of the mean number of absorbers $\bar{N}$. Individual absorbers have optical depth $\tau_0=1$. As in Figure \ref{fig1}, the 
Poisson limit is shown with the blue solid curve, while the highly clustered distributions for $b=0.2, 0.4, 0.6$ are shown with the 
red dot-dashed, dashed, and solid curves, respectively. 
}
\label{fig2}
\end{figure}

At a fixed $\bar{N}$, the clustering of absorption systems decreases
the opacity of the IGM compared to a random distribution, when viewed
from a random position.
This is shown in Figure \ref{fig2}, where we have used the probability
function in Equation (\ref{eq:nP}) to compute the effective optical
depth (in Euclidean space) for optically thick absorbers ($\tau_0=1$) at
varying ${\bar N}$ and clustering parameter $b$.  
This toy model shows how even moderate clustering ($b=0.2-0.4$) 
could result in a reduction of the effective absorption opacity of
15-45\%, greatly easing the tension between the directly measured 
``true" MFP and the one incorrectly inferred from \fnhi\ under 
the assumption of a randomly distributed population of thick absorbers.

\section{Summary and Future Work} 
\label{sec:summary}

In this manuscript, we have examined the principal observational
constraints on characterizing the \ion{H}{I} opacity of the IGM.  We
studied the tension between estimations of the mean free path \lmfp\
(Figures~\ref{fig:compare_mfp},\ref{fig:wfc3}; O13, R13), and
emphasized that current measurements of \fnhi\ at $\mnhi \approx
10^{15}-10^{17} \cm{-2}$ are in strong conflict
(Figure~\ref{fig:r13_k13}; R13, K13).  While some of these
disagreements may be the result of statistical variance, we argued
that they result primarily from two effects related to the clustering
of strong absorption-line systems.

The first effect is known as line-blending, the presence of two or
more absorption systems with comparable \nhi\ at small velocity
separation. 
Although line-blending has previously been recognized, its impact on
measurements of \fnhi\ and \lmfp\ have been under appreciated.
Going forward, it will be necessary to define \fnhi\ within 
well-defined velocity windows.  In particular, attempts to combine
\fnhi\ measurements from line-profile fitting with the observed
incidence of LLS have led to the double counting of Lyman limit opacity.
Specifically, to combine surveys of the LLS with measurements of
\fnhi\ from line-profile fitting one must `smooth' the latter by a
velocity window $\Delta v$.  Current LLS surveys based on lower resolution
spectra require $\Delta v \approx 2000\mkms$ \citep{pow10}.  
The size of $\Delta v$ may only be minimized through surveys at very
high spectral resolution and S/N covering the full Lyman series (e.g.\
the ESO X-Shooter Large Program; PI: Lopez).  

The other effect, the large-scale clustering of absorption systems, will likely require
calibration from cosmological simulations using radiative transfer.
It is also possible that one may introduce a clustering formalism
akin to the halo occupation distribution function for galaxies
\citep{tc08}.  Indeed, Zhu et al.\ (2013) have presented evidence for two
terms for the clustering of \ion{Mg}{II} systems around luminous red
galaxies, but the clustering of LLS is well modeled by a single
power-law \citep{qpq6}.  
To date, the LLS have been correlated with luminous, $z\sim 2$
quasars \citep{qpq2,qpq6}.  Further studies should examine the
auto-correlation function \citep{fumagalli13b} as well as the
cross-correlation function with the quasi-linear \lya\ forest.
And, ultimately, one must modify the definition for the effective
opacity of the IGM, possibly in
a manner similar to the toy model of $\S$~\ref{sec:toy}.

Returning to the MFP, there is yet another aspect of clustering
which may influence this measurement and one's estimate
for the intensity of the EUVB: the absorption systems are clustered
around the ionizing sources (quasars, galaxies).  \cite{qpq6} demonstrated that 
quasars are strongly clustered to optically thick gas, exhibiting a
covering fraction $f_C$ that approaches unity as one tends to small
impact parameters transverse to the sightline.  Extrapolating their
results to zero impact parameter (i.e.\ along the sightline or
`down-the-barrel'), one recovers $f_C > 0.8$.
This suggests that the ionizing radiation
field from quasars could be strongly attenuated.  One observes,
however, that very few quasars exhibit strong LL opacity at $z \approx
\mzq$ \citep{pow10}.  In fact, \cite{pow10} measured a {\it deficit}
of LLS within $\delta v = 3000 \mkms$ of $z \sim 3.7$ quasars relative to the
incidence measured at large velocity separations along the same
sightlines.  The natural interpretation is that quasars photoionize
the gas along the sightline, to distances of tens of Mpc
\citep[e.g.][]{qpq2}.
This quasar proximity effect may further increase \lmfp\ and the
resultant metagalactic flux. 
We encourage large volume simulations to explore these effects.

Of course, galaxies may also contribute to the EUVB, especially at
redshifts $z>4$ where one observes a steep decline in the comoving
number density of bright quasars \citep{fck06}.  Similar to the
quasar-LLS clustering, \cite{rudie12} have reported on an excess of
strong \ion{H}{I} absorption systems in the environment of Lyman break
galaxies (LBGs).  R13 further posited that the MFP from LBGs will be
smaller due to such clustering, although those authors ignored any
proximity effect associated to the LBG radiation field.  One can test
these effects by generating a composite spectrum in the LBG
rest-frame, akin to our quasar analysis.  We expect the data already
exist and encourage such analysis. 


\section*{Acknowledgments}

This research has made use of the Keck Observatory Archive (KOA),
which is operated by the W. M. Keck Observatory and the NASA Exoplanet
Science Institute (NExScI), under contract with the National
Aeronautics and Space Administration.
JXP acknowledges support from the National
Science Foundation (NSF) grant AST-1010004. 
P.M. acknowledges support  from the NSF through grant OIA-1124453, and
from NASA through grant NNX12AF87G.
Support for M.F. was provided by NASA through Hubble Fellowship grant
HF-51305.01-A awarded by the Space Telescope Science Institute, which
is operated by the Association of Universities for Research in
Astronomy, Inc., for NASA, under contract NAS 5-26555.
JMO acknowledges travel support from the VPAA's office at Saint
Michael's College.
We thank R. Cooke for help with MCMC analysis and with the use of his
ALIS software package.
We thank G. Rudie for providing a table of her \fnhi\ measurements and
Francesco Haardt for many useful conversations on the opacity of a
clustered IGM.


\end{document}